\documentclass[12pt,a4paper,onecolumn]{article}

\usepackage[T1]{fontenc}
\usepackage[utf8]{inputenc}
\usepackage{graphicx}
\usepackage{authblk}
\usepackage{xcolor}

\title{Formation energy of intrinsic defects in silicon from the Galitskii-Migdal formula.}
\author[1,2,3]{N.L. Matsko}
\affil[1]{\footnotesize Laboratory of Theoretical Physics, Joint Institute for Nuclear Research, 141980 Dubna, Russian Federation}
\affil[2]{\footnotesize Center for Photonics and 2D Materials, Moscow Institute of Physics and Technology, Dolgoprudny, Russian Federation}
\affil[3]{\footnotesize P.N. Lebedev Physical Institute, Russian Academy of Sciences, Leninskii prosp. 53, 119991 Moscow, Russian Federation}
\date{}

\begin{document}

\maketitle

\begin{abstract}

The work is devoted to the formation energy calculations of intrinsic defects in silicon based on the GW method and the Galitskii-Migdal formula. The two methods for calculating the electronic response function are applied. The first one uses direct integration over frequency to determine the response function. The diagonal form of the spectral function is the only assumption within the RPA framework, but the supercell calculations are very time-consuming. Therefore, we propose the method in which the response function is calculated in the plasmon pole approximation, and the GW contribution to the exchange-correlation energy is taken with a certain mixing constant. The value of the constant is found from the correspondence with experimental data. This makes it possible to obtain an accuracy comparable to the first method at significantly lower computational costs. The described method is used to calculate the formation energy of the neutral self-interstitial, vacancy and two divacansy structures in supercells of 214-217 silicon atoms.

\end{abstract}

\section{Introduction.}

The study of intrinsic defects in a silicon crystal is relevant because of the exceptional role of silicon devices in microelectronics. Such defects (vacancies, interstitials, etc.) arise under the influence of ionizing radiation, ion implantation, plasma etching, mechanical impact and other factors. Knowledge of the defect formation energy is necessary for calculating the concentration of defects, their movement through the crystal and finding the corresponding properties of the system. Experimental methods, however, face difficulties. The results are obtained by indirect methods, often differ markedly and can be interpreted in different ways \cite{dannefaer,bracht,bracht1995,ural,seebauer}. Therefore, theoretical methods are of great importance.

The defect formation energy $E_f$ is determined as follows. When removing k atoms from a certain region of a supercell with N atoms, we can write:
\begin{equation} \label{f_def0} E_f=E_{tot}(Si_{N})-E_{tot}Si_{N-k} + k\cdot\mu \end{equation}
where $E_{tot}(Si_{N})$ is the energy of an ideal supercell of N silicon atoms, $E_{tot}(Si_{N-k})$ is the energy of a supercell with a defect of N-k silicon atoms, $\mu=E_{tot}(Si_{N})/N$ is the chemical potential of silicon atom. Eventually:
\begin{equation} \label{f_def1} E_f=E_{tot}(Si_{N-k})-\frac{N-k}{N}E_{tot}(Si_{N}) \end{equation}

A feature of the defects under study is the appearance of electronic bands in the band gap of the crystal. It is well known that methods based on DFT and local-density functionals poorly describe such systems, greatly underestimating the size of the band gap, delocalizing defective electronic levels (electron self-interaction error). The defect formation energy in DFT is also underestimated. The reason for such errors lies in the consideration of the exchange-correlation (xc) contribution. For a correct quantitative description of a defect in a crystal, one needs to adequately account for the xc contribution, which will determine the accuracy of the approximation used. In general, the electronic response function could be expressed as the Dyson equation:

\begin{equation} \label{f_0} \chi(k\omega)=\chi_0(k\omega)+\chi_0(k\omega)\left[v(k)+f_{xc}(k\omega)\right]\chi(k\omega) \end{equation}

where $\chi_0(k\omega)$ is the response function of noninteracting particles, $v(k)$ is the Coulomb interaction, $f_{xc}(k\omega)$ is the xc kernel ($f_{xc}=\delta V_{xc}/\delta\rho$, $V_{xc}$ is the exchange-correlation functional). In the simplest case $f_{xc}=0$, which corresponds to the random phase approximation (RPA) \cite{pines}. Despite the simplicity, this approach makes it possible to describe screening and improve the treating of quasiparticle properties in semiconductors \cite{hyb-Lou}. Consideration of more complex kernels $f_{xc}$ greatly complicates the calculation, but does not always lead to an increase in accuracy \cite{fuchs}. At the moment, the use of such cores is limited to simple systems, such as electronic gas \cite{lein,jung,gg}, as well as systems containing dozens of atoms \cite{olsen13,olsen}.

The use of RPA to calculate the total energy of a system has a long history. Some papers claimed a noticeable increase in accuracy compared to the initial approximation. In others the result was the opposite. In the papers \cite{fuchs,olsen13,furche,harlkresse2010} the atomization energies of main-group molecules calculated with RPA were often worse than in DFT. In contrast, RPA describes well structural properties and binding energies of systems in which the vDW interaction plays a significant role \cite{harlkresse2010,lebharlkresse2010,luli,renrinke,barth,harris,pitarke,dobson}. 
This can be explained as follows. RPA is known to describe long-range correlations well. Accordingly, in systems where such a contribution is large, the results show an increase in accuracy. On the other hand, RPA poorely descripts short-range correlations, which leads, for example, to inaccurate account for covalent bond strengths, etc.
Thus, RPA is not suitable as a universal and highly precise post-DFT method for finding system energy  from the first principles. However, this is the method that goes beyond the semilocal approximation, strictly describes metal-type screening and shows how correlations affect the properties of the system \cite{harlkresse2010}. It makes RPA one of several major post-DFT approaches.

There are several ways to find the total energy within RPA. The RPA correlation energy can be calculated with the Adiabatic Connection Fluctuation Dissipation Theorem (ACFDT) by integrating over the coupling constant \cite{olsen13,kresse,niquet}. At the moment, it is one of the most commonly used approaches for calculating RPA total energy for realistic systems \cite{renrinke}. The total energy can be found variationally by functionals of the Green's function, such as the Klein and Luttinger-Ward functionals \cite{lw,lw1}. The shortcoming of this approach is that the functionals have an implicit form and are very complicated when calculating real systems.

In this paper, the system energy within the RPA framework is found by the Galitskii-Migdal formula \cite{gm} on the basis of a one-iterative GW calculation ($G_0W_0$) with the PBE initial approximation. The advantage of this approach is the simplicity of the Galitskii-Migdal analytical expression and the practical implementation of the GW method in many software packages.
GW is a perturbative method and initial approximation can significantly affect the results \cite{holm2}. This problem is removed in a self-consistent GW (sc-GW) that is norm-conserving, starting point independent approximation \cite{baym1,baym2}. For electron gas sc-GW gives total energy in perfect agreement with Monte Carlo calculations \cite{ceperley,gg1}. But even for such a system, this is a computationally expensive task. In recent years, sc-GW calculations for real systems have also appeared \cite{delaney,dahlen,kutepov}. The accuracy of sc-GW total energy calculation decreases as the behavior of the system moves away from the case of an electron gas \cite{dvb1,dvb2,dvb3}. In addition, self-consistency in GW can lead to negative effects in the description of the quasi-particle spectrum: excessive broadening of the valence band, the appearance of non-physical features in the spectral function, distortion of the satellite structure \cite{holm}. There are no such problems in $G_0W_0$, which is due to the fact that errors from the lack of self-consistency and the absence of vertex corrections are mutually subtracted.
$G_0W_0$ is accurate for semiconductors \cite{olsen13,kresse} and is widely used in practice. Moreover, crystalline silicon is a reference example for $G_0W_0$ PBE calculation.
Thus, the approximation used should be accurate for systems of silicon defects too.
In addition to the calculations of the defect formation energies, the effectiveness of the proposed approach will be analyzed, since its accuracy is limited by taking into account only RPA-type correlations, while calculations can be quite time-consuming.

\section{Theory.}

The total energy is given by:
\begin{equation} \label{f_2} E_{tot}\{\rho\}=E_{kin}+\int V_{ext}(r)\rho(r)d^3r+\frac12\int d^3rd^3r' \hat\psi^\dag(r) \hat\psi^\dag(r')v(r-r')\hat\psi(r) \hat\psi(r') \end{equation}

where the first term is kinetic energy, the second is the energy in an external field, the last term describes the electron-electron interaction. The equation of motion for the Green's function $G$ is:

\begin{equation} \label{f_3} \left[ i\frac{\partial}{\partial t} - h(r) \right]G(rt,r't') - \int d^3rd^3r' \hat\psi^\dag(r) \hat\psi^\dag(r')v(r-r')\hat\psi(r) \hat\psi(r')= \delta(r-r') \end{equation}

where $ h(r)=-\frac{\hbar^2}{2m}\nabla^2+V_{ext}(r)$ is the single-particle operator. After expressing the last term in the left part of (\ref{f_3}) (electron-electron interaction) and substituting into (\ref{f_2}), we get:

\begin{equation} \label{f_4} E_{tot}\{\rho\}= \frac12\int d^3r\left[  i\frac{\partial}{\partial t} + h(r) \right] G(rt,r't')_{r'\to r,t'\to t^{\dag}}  \end{equation}

Expression (\ref{f_4}) is the Galitskii-Migdal formula for the total energy depending on the single-particle Green's function. Spectral representation for Green's function:

\begin{equation} \label{f_5} G(r,r';\omega)= \int\limits_{-\infty}^{\mu} d\omega'\frac{A(r,r';\omega)}{\omega-\omega'-i\delta} + \int\limits^{\infty}_{\mu} d\omega'\frac{A(r,r';\omega)}{\omega-\omega'+i\delta} \end{equation}

After substituting the spectral representation in (\ref{f_4}):

\begin{equation} \label{f_6} E_{tot}\{\rho\}= \frac12\int drd\omega \left[ \omega + h(r) \right] A(r,r';\omega)_{r'\to r}  \end{equation}

We express the spectral density in the Kohn-Sham basis:
\begin{equation} \label{f_6.1} A(r,r';\omega)= \sum_{knn'}\psi_{kn}(r)A_{nn'}(k,\omega)\psi^*_{kn'}(r') \end{equation}
and substitute in (\ref{f_6}). Substituting also $ h(r)$ from the Kohn-Sham equation $[ h+ V_H+ V_{xc}]\psi_{kn}=\epsilon_{kn}\psi_{kn}$ and assuming the spectral function to be diagonal  $\int\limits_{-\infty}^{\mu}d\omega A_{nn'}(k,\omega)=n_{kn}\delta_{nn'}$ we get:

\begin{equation} \label{f_7} E_{tot}\{\rho\}= \frac12\int d\omega \omega A(\omega)+\sum\limits_{kn}n_{kn}\epsilon_{kn} - \int dr V_H\rho(r) - \int dr V_{xc}\rho(r) \end{equation}

Such an approximation is valid, since calculations show that the non-diagonal elements are very small \cite{hyb-Lou,holm2, arya1992}. When considering only the diagonal self-energy elements in silicon, it is possible to obtain quasi-particle energies with an error of less than 0.05 eV \cite{hyb-Lou}. Moreover, off-diagonal elements can change sign, so the integration over frequency makes their contribution even less \cite{holm2,hedin}.

The DFT total energy is:

\begin{equation} \label{f_8} E_{tot}^{DFT}= E_{kin} + \int V_{ext}(r)\rho(r)dr + \frac{e^2}{2}\int \frac{\rho(r)\rho(r')}{|r-r'|}drdr' + \int \epsilon_{xc}\{ \rho(r)\} \rho(r) dr \end{equation}

Let's express $E_{kin}$ from the Kohn-Sham equation:

\begin{equation} \label{f_9} E_{kin}=-\sum_{kn}\frac{\hbar^2 \nabla^2}{2m}\psi_{kn}^2=\sum_{kn} \epsilon_{kn} n_{kn} - \sum_{kn}\int V_{eff}(r) \psi_{kn}^2(r) dr= \end{equation}

\begin{equation} \label{f_10} = \sum_{kn} \epsilon_{kn} n_{kn} - \int V_{ext}(r) \rho(r) dr - e^2\int \frac{\rho(r)\rho(r')}{|r-r'|}drdr' - \int V_{xc}(r) \rho(r) dr \end{equation}

after substitution into (\ref{f_8}) we get:

\begin{equation} \label{f_11} E_{tot}^{DFT}= \sum\limits_{kn}n_{kn}\epsilon_{kn} - \frac{e^2}{2}\int \frac{\rho(r)\rho(r')}{|r-r'|}drdr'+ \int \epsilon_{xc}\{ \rho(r)\} \rho(r) dr - \int dr V_{xc}(r)\rho(r) \end{equation}

together with (\ref{f_7}) we obtain the Galitskii-Migdal correction to the DFT total energy \cite{holm2}:

\begin{equation} \label{f_12} E_{tot}-E_{tot}^{DFT}=- E_{xc} + \frac12\left[ \int\limits_{-\infty}^{\mu}d\omega \omega A(\omega)-\sum\limits_{i}n_i\epsilon_{i}^{DFT} +  \sum\limits_{i} V_{i}^{xc}\right]  \end{equation}

where $V_{i}^{xc}=<\psi_i|\hat V_{xc}|\psi_i>$. The formula (\ref{f_12}) will be used further to calculate the total energy.

When calculating with the use of generalized plasmon pole approximation (GPP approximation), the spectral function was considered as a set of delta functions at quasiparticle energies $\epsilon_{i}^{GW}=\epsilon_{i}^{DFT} - V^{xc}_i+\Sigma^{GPP}_i$ \cite{deslippe}. Then:

\begin{equation} \label{f_13} E_{tot}^{GW}= E_{tot}^{DFT}- E_{xc} + \frac12\left[ \sum\limits_{i}(\epsilon_{i}^{GW}-\epsilon_{i}^{DFT})n_i +  \sum\limits_{i} V_{i}^{xc}\right]= E_{tot}^{DFT}- E_{xc} + \frac12\sum\limits_{i} \Sigma_i^{GPP}  \end{equation}

Simplification in the form of GPP makes the resulting correction to energy rougher and overestimated. So it's logical to take it into account with some weight factor. This approach is  similar to the hybrid functional method \cite{hybr1,hybr2,hybr3} - when the exchange-correlation energy includes the exact exchange multiplied by the mixing constant $\alpha$, the rest is calculated by the corresponding DFT functional. In our case, not only static correlations at the Hartree–Fock level are taken into account, but also dynamic correlations at the RPA level:

\begin{equation} \label{f_16} E_{tot}^{GPP}= E_{tot}^{DFT} - \alpha \cdot E_{xc}^{DFT} + \alpha \cdot \Sigma^{GPP}= E_{tot}^{DFT}-\alpha \cdot E_{xc}^{DFT} + \alpha \cdot \frac12 \sum\limits_{i}\left[ \epsilon_{i}^{GW}-\epsilon_{i}^{DFT} + V_{i}^{xc}\right] \end{equation}

\section{Calculation details.}

DFT calculations in the Quantum Espresso code \cite{qe} with PBE GGA pseudopotential were used as a starting point for the GW computations. The calculations in crystalline silicon were carried out for the diamond structure with a cubic unit cell edge of 5.43 \AA, plane wave basis cutoff energy was set to 45 Ry. 
Calculations in 64-65 atom supercells were carried out on a uniform grid containing 3*3*3 k-points in the case of the full frequency dielectric matrix calculations, and containing 4*4*4 k-points in the case of the GPP approximation. The total energy for a given structure was found as a weighted average for three k-points ($\Gamma$; 0,0,1/3; 0,1/3,2/3), with k-point weight corresponding to its symmetry degeneration (1, 3 and 6 correspondingly).
Calculations in 214-217 atom supercells were carried out on a uniform grid containing 2*2*2 k-points. The total energy was calculated at the $\Gamma$ point, at other k-points the total energy differed by no more than 0.01 eV.

For GWA calculations the BerkeleyGW \cite{hyb-Lou,deslippe,bgw2} package was applied. The unoccupied bands were calculated up to 20 eV above the Fermi surface. Energy cutoff for the dielectric matrix was set to 5 Ry. The two methods for calculating the total energy from GW were used. In the first method (full-frequency GW), the dielectric response function is found by integrating over frequency with the contour-deformation formalism. The dielectric matrix is evaluated on both real and imaginary frequencies: the uniform grid with the step of 0.25 eV for the frequency range of 0-46 eV on the real axis plus 15 imaginary frequencies. The spectral function is integrated over the frequency interval starting from 50 eV below the Fermi energy and ending at an energy 1 eV above the Fermi energy. Such an interval ensures the integration of all features of the spectral function, taking into account its satellite structure \cite{matsko18}.The total energy for the chosen k-point was calculated by the formula (\ref{f_12}).
It is worth noting that the calculation by the full-frequency GW for 64-65 atom supercells turns out to be computationally extremely expensive. The calculation of one k-point took 3-4 days for 2048 Intel Xeon E5-2697Av4 cores.

The second method of calculating the total energy uses the GPP approximation to find the dielectric matrix. The computational requirements in this case are an order of magnitude less in comparison with full-frequency GW. The total energy is found by the formula (\ref{f_16}) based on the calculated quasiparticle energies $\epsilon_{i}^{GW}=\epsilon_{i}^{DFT}+<\psi_i|\Sigma(\epsilon_{i}^{GW})-V_{xc}|\psi_i>$. The energy $\epsilon_{i}^{GW}$ appears in the right and left sides of the equation. To find a self-consistent value we make iterations of the substitutions $\epsilon_{i}^{GW}$ into $\Sigma$ for next step until convergence is achieved (self-consistency in the eigenvalues or ev-scGW) \cite{hyb-Lou}. This reduces the dependence of the GW calculation on the initial approximation \cite{marom} and, in most cases, increases the accuracy of the quasiparticle energies \cite{marom,blase,faber}. Comparison of GPP GW results with experimental data, as will be shown below for several calculations, demonstrates that the value of the optimal mixing constant $\alpha$ belongs to the range of 5\%-15\%. Note, for example, that in the hybrid functional PBE0, the mixing constant is higher and usually amounts to about 20\%-25\% \cite{hybr2,hybr3}. By mentioning the GW calculation below, we will mean a one-iterative calculation of G$_0$W$_0$.

The calculations of the energy for silicon cluster dissociation $Si_4\rightarrow2Si_2$ and $Si_6\rightarrow2Si_3$ were carried out at the $\Gamma$ point with the same parameters as calculations in a silicon crystal. A cubic supercell with a side of 50 Bohr was used. This size provides a sufficient vacuum layer to minimize the impact of replicas from neighboring supercells. The dissociation energy was defined as the difference between the total energy of the initial cluster and the energy when the cluster is divided into two equal parts separated by a distance of half the diagonal of the supercell. In the $Si_2$ cluster, the PBE gives fractional occupancies of the electronic levels. Therefore, when calculating the total energy with formulas (\ref{f_12}) and (\ref{f_16}), the corresponding occupation numbers should be substituted.

The structures of the considered defects were relaxed using PBE GGA functional until atomic forces became less than $10^{-4}$ Ry/\AA. Si vacancy in a neutral charge state has the symmetry D$_{2d}$. Self-interstitial has several structures that are close in energy. We consider the most stable structure split <110> \cite{rinke}, the geometry of which is shown in Fig. 1. The geometries of the two most energetically favorable divacancies \cite{jay}, which were considered in the paper, are also shown in Fig. 1. The defect formation energy was determined by the formula (\ref{f_def1}).

\begin{figure}[h]
\centering
\includegraphics[width=0.8\textwidth]{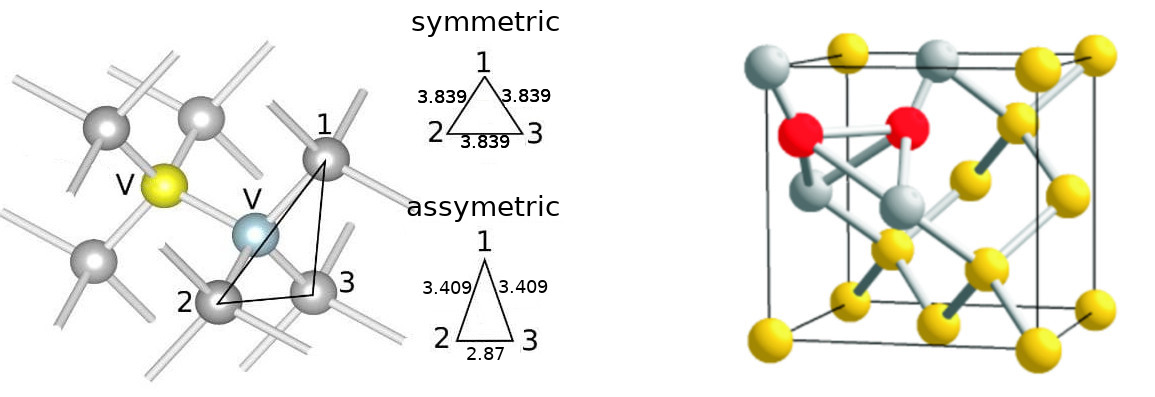}
\caption{Structures of symmetric and asymmetric divacancies (left part), self-interstitial structure with the split <110> symmetry (right part). Length are given in angstroms.}
\end{figure}

\section{Results and discussion}

To estimate the methods used, we have calculated the energy for cluster dissociation $Si_4\rightarrow2Si_2$ and $Si_6\rightarrow2Si_3$. These examples are indicative as the dangling bonds of silicon affect the energy of clusters, as in the case of intrinsic defects in the crystal. However, the screening in such small objects differs from the screening in a crystal and is described worse in terms of RPA \cite{furche}.
Table 1 shows the experimental values for the $Si_4$ and $Si_6$ dissociation energies, the results in   PBE, hybrid functionals PBE0, B3LYP and HSE06, as well as in the full-frequency GW and GPP GW with mixing of 15\%, 10\% and 5\%. In the case of $Si_4\rightarrow2Si_2$ the dissociation energy error is 0.98 eV (17.5\%) for PBE and 1.02 eV (18\%) for PBE0, 0.3 eV (5\%) for B3LYP, 2 eV (36\%) for HSE06, 0.26 eV (4.5\%) for full-frequency GW. The dissociation energy error of $Si_6\rightarrow2Si_3$ is 0.05 eV (1\%) for PBE and 0.27 eV (4.5\%) for PBE0, -1.03 eV (18\%) for B3LYP, 0.26 eV (4.5\%) for HSE06, 0.01 eV (0.2\%) for full-frequency GW. The accuracy of PBE and hybrid functionals is noticeably lower than the accuracy of full-frequency GW.

When calculating the $Si_4$ dissociation energy with GPP GW, the optimal mixing value is about 15\%. In the case of $Si_6$ dissociation, the optimal mixing is slightly less than 5\%. This difference is explained by the fact that the examples considered represent two extreme cases. Within the framework of PBE, the $Si_6\rightarrow2Si_3$ dissociation is described well, while the $Si_4\rightarrow2Si_2$ dissociation energy exhibits a serious error, which is caused by an inaccurate description of the $Si_2$ dimer. In $Si_2$, only one electron per atom participates in the formation of a covalent bond, the other 3 valence electrons remain unpaired, delocalized and make a large contribution to the electron self-interaction error. In the $Si_3$ cluster (obtained in $Si_6$ dissociation), the self-interaction error is noticeably smaller, since 2 unpaired electrons on each atom are localized on opposite sides of the cluster plane and interact more weakly.
These two considered cases allow us to estimate the limits of the mixing constant in GPP GW, a typical value should be in this range. In DFT PBE calculations for a silicon crystal, the defective bands depend  on the self-interaction error noticeably. Therefore, for GPP GW it is logical to expect the mixing in the region of 15\%-10\%, comparable to mixing for $Si_4\rightarrow2Si_2$ dissociation.

\begin{tabular}{|p{3.0cm}|p{5.9cm}|p{5.95cm}|}
\multicolumn{3}{l}{}\\
\multicolumn{3}{l}{\textbf{Table 1.} Dissociation energy of $Si_4$ and $Si_6$ clusters. For the}\\
\multicolumn{3}{l}{GPP G$_0$W$_0$ the mixing constant is given in brackets.}\\
\hline   & $Si_4\rightarrow2Si_2$ diss. energy, eV & $Si_6\rightarrow2Si_3$  diss. energy, eV  \\
\hline             Experiment \cite{dissoc} & 5.6 & 5.82  \\
\hline full-freq. G$_0$W$_0$ & 5.86 & 5.83 \\
\hline GPP G$_0$W$_0$  & 5.62(15\%), 5.94(10\%), 6.26 (5\%) & 5.64 (15\%), 5.71 (10\%), 5.79 (5\%)\\
\hline    PBE     & 6.58 & 5.87  \\
\hline    PBE0    & 6.62 & 6.09 \\
\hline    B3LYP   & 5.9 & 4.79  \\
\hline    HSE06     & 7.6 & 6.08  \\
\hline
\multicolumn{3}{l}{}
\end{tabular}

\begin{tabular}{|p{3.0cm}|p{2.5cm}|p{2.5cm}|p{2.8cm}|p{2.5cm}|}
\multicolumn{5}{l}{}\\
\multicolumn{5}{l}{\textbf{Table 2.} Defect formation energy $E_f$ in a supercell, eV. For the}\\
\multicolumn{5}{l}{GPP G$_0$W$_0$ the mixing constant is given in brackets.}\\
\hline \multicolumn{5}{|c|}{64-65 atom supercell}\\
\hline   &  vacancy &  interstitial split <110> & divacansy nonsymmetric & divacancy symmetric  \\
\hline PBE & 3.6 & 3.62 &  & 5.4 \\
\hline full-freq. G$_0$W$_0$ & & 6.07 &  &  \\
\hline GPP G$_0$W$_0$ &  & 5.03 (15\%) \quad {\bf4.56} (10\%) \quad 4.09 \quad (5\%) & & \\
\hline \multicolumn{5}{|c|}{214-217 atom supercell}\\
\hline   &  vacancy &  interstitial split <110> & divacansy nonsymmetric & divacancy symmetric  \\
\hline PBE & 3.63 & 3.66 & 5.1 & 5.19 \\
\hline GPP G$_0$W$_0$ & 6.05 (15\%) \quad {\bf5.24} (10\%) \quad 4.44 \quad (5\%)  & 5.03 (15\%) \quad {\bf4.58} (10\%) \quad 4.12 \quad (5\%) & 6.21 (15\%) \quad {\bf5.84} (10\%) \quad 5.47 \quad (5\%) & 6.32 (15\%) \quad {\bf5.95} (10\%) \quad 5.57 \quad (5\%) \\
\hline
\multicolumn{5}{l}{}
\end{tabular}

The upper part of Table 2 contains the results of the $E_f$ calculations for self-interstitial in a 65 atom supercell by the PBE, full-frequency GW, and GPP GW methods. As a reference value for the self-interstitial $E_f$, we consider the range of 5-4.4 eV, which is based on quantum Monte Carlo calculations: 4.96 eV in \cite{leung}, 4.94 eV in \cite{batista}, and also 4.4 eV in \cite{parker}. Experimental energy values vary too much to be reference ones. For example, the experimental activation energy (formation energy + migration barrier) of an interstitial was determined as 4.68 eV in \cite{ural} and as 5.8 eV in \cite{ghaderi}. The migration barrier values are determined from 0.13 to 2 eV \cite{kato1993}.
The full-frequency GW in a 65 atom supercell gives an interstitial formation energy of 6.07 eV. This overestimates the reference value by 1-1.5 eV (while PBE underestimates by the same amount). 
Apparently, a significant error of full-frequency GW is explained by insufficient calculation parameters. A more accurate description of the spectral function in the region of the Fermi surface (where it has a form close to $\delta$ function) is needed. The GPP GW with 10\% mixing gives an interstitial $E_f$ of 4.56 eV, which lies in the reference range. Further, 10\% mixing for GPP GW will be considered as the default, the corresponding values in the table are marked in bold.

The second part of Table 2 contains the results of the $E_f$ calculations in a 214-217 atom supercell by the PBE and GPP GW methods for vacancy, interstitial and two divacancies. The use of full-frequency GW in such a supercell requires excessive computing resources. The interstitial formation energy in GPP GW is 4.58 eV and coincides with the value for a 65 atom supercell, which indicates that convergence in the supercell parameter for the interstitial has been achieved. The interstitial has the lowest $E_f$. 
The structures next on the energy scale are a vacancy with $E_f$ = 5.23 eV, and two divacancies with $E_f$ = 5.84 eV and $E_f$ = 5.95 eV.
This ordering differs from PBE calculations, in which the interstitial and vacancy have approximately equal $E_f$, that is 1.5 eV less than $E_f$ for divacancies. In all considered defects, PBE shows a noticeably lower formation energy in comparison with GW. Underestimation of $E_f$ is typical in local-density functionals, including defects in other materials \cite{thomas,ryu}. GW corrects the PBE formation energy towards the right value. This correction is usually excessive due to the overestimation of the correlation energy contribution in RPA \cite{olsen13,olsen,harlkresse2010}.

For comparison, we present the $E_f$ obtained using RPA by other authors for the split <110> interstitial, vacancy and divacancy. In \cite{rinke} interstitial $E_f$ equals 4.46 eV. In \cite{kaltak} $E_f$ equals 4.33 eV for a vacancy and 4.20 eV for an interstitial. In \cite{bruneval} the range-separated RPA gives $E_f$ of a vacancy equal to 4.33 eV and to 4.49 eV for an interstitial. In \cite{yao} the RPA with the exact exchange included gives $E_f$ equal to 4.51 eV for a vacancy, 4.24 eV for an interstitial and 6.59 eV for a divacancy. The range-separated RPA gives $E_f$ equal to 4.26 eV for a vacancy, 4.42 eV for an interstitial, 6.79 eV for a divacancy in \cite{yao}. The structure of the divacancy in \cite{yao} is not specified.
As it can be seen the $E_f$ for each defect vary by 0.2-0.3 eV for different works. The vacancy formation energy in given works is almost 1 eV higher than our value. The interstitial $E_f$ in our calculations is in good agreement with the results of \cite{rinke,bruneval}, as well as the range-separated RPA in \cite{yao}. The $E_f$ for divacancies in our calculations are almost 1 eV lower than in \cite{yao}, however the divacancy structure considered in \cite{yao} can be in principle higher in energy.

\section{Conclusions}

Total energy calculations in the framework of Galitsky-Migdal formalism based on the G$_0$W$_0$ method were made. It was shown that in silicon structures with dangling electronic bonds this approach significantly improves the accuracy of the total energy differences compared to initial PBE approximation. The approach was applied to the formation energy calculations of the neutral intrinsic defects in a silicon supercell.
The full-frequency GW calculations for 64-65 atom supercell turn out to be very time consuming, while the value of self-interstitial $E_f$ is greatly overestimated. Improving of the accuracy requires better calculation parameters.
To circumvent the limitation of excessive computational requirements, a method has been proposed where  the response function was calculated in the plasmon pole approximation. In this case, the value of the exchange-correlation energy is usually overestimated compared to the full-frequency GW. Therefore, the contribution of GW was taken with a certain mixing constant. The optimal mixing coefficient of 10\% was determined from the analysis of GPP GW results for $Si_4$ and $Si_6$ clusters dissociation and the calculation of $E_f$ for the silicon self-interstitial, . The demonstrated accuracy becomes comparable to full-frequency GW, while the computational requirements are much lower. Eventually this allows to calculate the defect formation energies in supercells of 214-217 silicon atoms. 
The following values were obtained: 4.58 eV for an interstitial, 5.23 eV for a vacancy, and 5.84 eV and 5.95 eV for asymmetric and symmetric divacancies. The applied approach is conceptually similar to the hybrid functional in DFT, but in comparison to the latter, it takes into account the effects of dynamic correlations beyond the local approximation.

\section{Acknowledgements}

Calculations were carried out on the Joint Supercomputer Center of the Russian Academy of Sciences (JSCC RAS).

\bibliography{thesis}
\bibliographystyle{gost705}

\end{document}